\documentclass[prc,preprint,showpacs,superscriptaddress,byrevtex]{revtex4}

\usepackage{epsfig}
\usepackage{amssymb}
\usepackage{graphicx,color}
\usepackage{dcolumn}

\begin{document}

\newcommand{\bra}[1]{\langle #1 \vert}
\newcommand{\ket}[1]{\vert #1 \rangle}
\newcommand{\rbra}[1]{\langle #1 \vert\vert}
\newcommand{\rket}[1]{\vert\vert #1 \rangle}
\newcommand{\expec}[1]{\langle #1 \rangle}
\newcommand{\braket}[2]{\langle #1 \vert #2 \rangle}
\def\npdg{$\overrightarrow{n} + p \rightarrow d + \gamma$}
\def\Al{${\rm ^{27}Al}$}
\def\Cu{${\rm ^{59}Cu}$}
\def\In{${\rm ^{115}In}$}
\def\CompAl{${\rm ^{28}Al}$}
\def\CompCu{${\rm ^{60}Cu}$}
\def\CompIn{${\rm ^{116}In}$}
\def\npdg{$\vec{n} + p \rightarrow d + \gamma$}
\def\BC{${\rm B_4C}$}
\def\CCL{${\rm CCl_4}$}
\def\REAC{$^{10}$B(n,$\alpha$)$^{7}$Li}
\def\LIP{${\rm ^6Li}$}
\def\Ag{$A_{\gamma}$}
\def\Alr{$A_{\gamma,LR}$}
\def\Gr{$\gamma$-ray}
\def\Grs{$\gamma$-rays}
\def\Hpi{$H_{\pi}$}

\preprint{LAUR-XXXX-05}

\title{Upper Bounds on Parity Violating Gamma-Ray Asymmetries in Compound Nuclei from Polarized Cold Neutron Capture.}

\author{M.T.~Gericke}
\email[Corresponding author. Tel.: + 1-757-269-7346 ]{ mgericke@jlab.org}
\affiliation{University of Manitoba, Winnipeg, MB, Canada R3T 2N2}
%\affiliation{TRIUMF, 4004 Wesbrook Mall, Vancouver, BC, Canada V6T 2A3}
\affiliation{Thomas Jefferson National Accelerator Facility, Newport News VA 23606, USA}
\author{ J.D.~Bowman}
%\thanks{Spokesperson}
\affiliation{Los Alamos National Laboratory, Los Alamos, NM 87545, USA}
\author{R.D.~Carlini}
\affiliation{Thomas Jefferson National Accelerator Facility, Newport News VA 23606, USA}
\author{T.E.~Chupp}
\affiliation{University of Michigan, Ann Arbor, MI 48104, USA}
\author{K.P.~Coulter}
\thanks{Present address: General Dynamics-Advanced Information Sys. 1200 Hall Drive, Ypsilanti, MI 48197, USA}
\affiliation{University of Michigan, Ann Arbor, MI 48104, USA}
\author{M.~Dabaghyan}
\affiliation{University of New Hampshire, Durham, NH 03824, USA}
\author{M.~Dawkins}
\affiliation{Indiana University, Bloomington, IN 47405, USA}
\author{D.~Desai} 
\thanks{Present address: Department of Radiation Medicine, University of Kentucky, Lexington, KY 40506, USA}
\affiliation{University of Tennessee, Knoxville, TN 37996, USA}
\author{S.J.~Freedman}
\affiliation{University of California, Berkeley, CA 94720-7300, USA}
\author{T.R.~Gentile}
\affiliation{National Institute of Standards and Technology,Gaithersburg, MD 20899-0001,USA}
\author{R.C.~Gillis}
\affiliation{University of Manitoba, Winnipeg, MB, Canada R3T 2N2}
\author{G.L.~Greene}
\affiliation{University of Tennessee, Knoxville, TN 37996, USA}
\affiliation{Oak Ridge National Laboratory, Oak Ridge, TN 37831, USA}
\author{F.W.~Hersman}
\affiliation{University of New Hampshire, Durham, NH 03824, USA}
\author{T.~Ino}
\affiliation{High Energy Accelerator Research Organization (KEK), Tukuba-shi, 305-0801, Japan}
\author{G.L.~Jones}
\affiliation{Hamilton College, Clinton, NY 13323, USA}
\author{M.~Kandes}
\affiliation{University of Michigan, Ann Arbor, MI 48104, USA}
\author{B.~Lauss}
\affiliation{University of California. Berkeley, CA 94720-7300, USA}
\author{M.~Leuschner}
\affiliation{Indiana University, Bloomington, IN 47405, USA}
\author{W.R.~Lozowski}
\affiliation{Indiana University, Bloomington, IN 47405, USA}
\author{R.~Mahurin}
\affiliation{University of Tennessee, Knoxville, TN 37996, USA}
\author{M.~Mason}
\affiliation{University of New Hampshire, Durham, NH 03824, USA}
\author{Y.~Masuda}
\affiliation{High Energy Accelerator Research Organization (KEK), Tukuba-shi, 305-0801, Japan}
\author{G.S.~Mitchell}
\thanks{Present address: Department of Biomedical Engineering, University of California, Davis, CA 95616, USA}
\affiliation{Los Alamos National Laboratory, Los Alamos, NM 87545, USA}
\author{S.~Muto}
\affiliation{High Energy Accelerator Research Organization (KEK), Tukuba-shi, 305-0801, Japan}
\author{H.~Nann}
\affiliation{Indiana University, Bloomington, IN 47405, USA}
\author{S.A.~Page}
\affiliation{University of Manitoba, Winnipeg, MB, Canada R3T 2N2}
\author{S.I.~Penttil{\"a}}
\thanks{Present Address: Oak Ridge National Laboratory, Oak Ridge, TN 37831}
\affiliation{Los Alamos National Laboratory, Los Alamos, NM 87545, USA}
\author{W.D.~Ramsay}
\affiliation{University of Manitoba, Winnipeg, MB, Canada R3T 2N2}
\affiliation{TRIUMF, 4004 Wesbrook Mall, Vancouver, BC, Canada V6T 2A3}
\author{S.~Santra} 
\thanks{Present address: Bhabha Atomic Research Center, Trombay, Mumbai 400085, India}
\affiliation{Indiana University, Bloomington, IN 47405, USA}
\author{P.-N.~Seo}
\thanks{Present address: Department of Physics, North Carolina State University, Raleigh, NC 27695, USA} 
\affiliation{Los Alamos National Laboratory, Los Alamos, NM 87545, USA}
\author{E.I.~Sharapov}
\affiliation{Joint Institute for Nuclear Research, Dubna, Russia}
\author{T.B.~Smith}
\affiliation{University of Dayton, Dayton, OH 45469, USA}
\author{W.M.~Snow}
\affiliation{Indiana University, Bloomington, IN 47405, USA}
\author{W.S.~Wilburn}
\affiliation{Los Alamos National Laboratory, Los Alamos, NM 87545, USA}
\author{V.Yuan}
\affiliation{Los Alamos National Laboratory, Los Alamos, NM 87545, USA}
\author{H.~Zhu}
\affiliation{University of New Hampshire, Durham, NH 03824, USA}
\collaboration{The NPDGamma Collaboration}

\date{\today}

\begin{abstract}
Parity-odd asymmetries in the electromagnetic decays of compound
nuclei can sometimes be amplified above values expected from simple
dimensional estimates by the complexity of compound nuclear states. In
this work we use a statistical approach to estimate the root mean
square (RMS) of the distribution of expected parity-odd correlations $\vec{s_{n}} \cdot
\vec{k_{\gamma}}$, where $\vec {s_{n}}$ is the neutron spin and
$\vec{k_{\gamma}}$ is the momentum of the gamma, in the integrated
gamma spectrum from the capture of cold polarized neutrons on Al, Cu,
and In and we present measurements of the asymmetries in these and
other nuclei. Based on our calculations, large enhancements of asymmetries
were not predicted for the studied nuclei and the statistical
estimates are consistent with our measured upper bounds on the
asymmetries.
 
\end{abstract}

%21.30.Cb,   % Nuclear forces in vacuum
%11.30.Er,   % Charge conjugation, parity, time reversal, and other
             % discrete symmetries.
%24,70.+s    % Polarization phenomena in reactions
%25.40.Cm    % Elastic proton scattering.
%13.75.Cs    % Nucleon-Nucleon Interactions
%07.85.-m    % X- and gamma-ray instruments
%25.40.Lw    % Radiative Capture

\maketitle

\section{Introduction}
\label{scn:INSU}
One might assume that a quantitative treatment of symmetry breaking in
neutron reactions with heavy nuclei would not be feasible. However,
theoretical approaches exist which exploit the large number of
essentially unknown coefficients in the Fock space expansion of
complicated compound nuclear states in heavy nuclei to perform
calculations that can be compared to experiment. If we assume that it
is possible to treat the Fock space components of the states as
independent random variables, one can devise statistical techniques to
calculate, not the value of a particular observable, but the root mean
square of the distribution of expected values. This strategy has been
used successfully to understand certain global features of nuclear
structure and reactions~\cite{bb:WONG}. The distribution of energy
spacings and neutron resonance widths, for example, has been known for
a long time to obey a Porter-Thomas distribution~\cite{pp:Porter65} in
agreement with the predictions of random matrix theory, and
statistical approaches have been used to understand isospin violation
in heavy nuclei~\cite{pp:Harney86}.

The complexity of the compound nuclear states can also amplify the
size of the parity-odd asymmetries by several orders of magnitude
relative to single-particle estimates. This large amplification makes
it practical to use nuclear parity violation, generically expected on
dimensional grounds to possess amplitudes seven orders of magnitude
smaller than strong interaction amplitudes, as a new setting to
investigate the validity of these statistically-based theoretical
approaches.  Statistical analyses have successfully been applied
recently to an extensive series of measurements of the parity-odd
correlation $\vec{s_{n}} \cdot {\vec p_{n}}$ in the A=100-200 mass
region in neutron-nucleus scattering performed at Dubna, KEK, and
LANSCE~\cite{pp:Mitchell99,pp:TRIPLE, pp:FLSU, pp:FLGR}. Although the
comparison between theory and experiment in this work is still
hampered somewhat by the lack of precise knowledge of the weak NN
amplitudes and their possible modifications in the nuclear medium,
theory and experiment appear to be in agreement at about the 50\%
level. Given the extreme complexity of the states involved, agreement
between theory and experiment at this level must be counted as an
overall success for the statistical approach.

Parity violation in the gamma decays of nuclei is another example
where statistical methods may be employed to estimate observables. In
this case the observables involve the parity-odd correlation
$\vec{s_{n}} \cdot \vec{k_{\gamma}}$ where $\vec {s_{n}}$ is the
neutron spin and $\vec{k_{\gamma}}$ is the momentum of the gamma
~\cite{pp:FLSU,pp:FLGR,pp:HAYES}. Just as for neutron scattering,
neutron capture on elements with a large number of nucleons produces
compound nuclei in highly excited states.  These nuclei exhibit a huge
$(>10^6)$ number of possible state configurations with different
angular momenta and parity and the number of transitions with
different amplitudes that the compound nucleus may make to its ground
state is correspondingly large as well. Because of the large number of
energy levels in the compound nucleus, formed by neutron capture, one
may hope that the calculation of the mean square matrix elements for
the transition amplitudes may also amount to a summation of a large
number of uncorrelated random contributions as in the case of the
total cross section.  One can then use statistical arguments to
estimate the RMS value of the parity-odd \Gr~asymmetry.

However the case of parity violation in (n,$\gamma$) reactions in
heavy nuclei is not quite as simple as parity violation in the total
cross section for both theoretical and experimental reasons. For the
total cross section, the amplification of parity violation effects is
dominated by the mixing amplitude of the weak interaction, between two
compound nuclear states of opposite parity (in practice s-wave and
p-wave compound states). Since the total cross section is proportional
to the forward elastic scattering amplitude, by the optical theorem,
there is only one such contribution for any pair of opposite-parity
compound states. For inelastic processes such as the (n,$\gamma$)
reaction, however, the weak mixing between compound states can occur
in either the initial or final nuclear states, and since these states
are distinct in an inelastic reaction there are two possible sources
of compound nuclear amplification of the parity-odd effect rather than
one~\cite{pp:FLGR}. Because of the large density of states in the
initial state near neutron separation energy, the initial state mixing
will involve a larger number of components in the wave function for
gamma transitions to low-lying states and therefore lead to a larger
amplification. However one also has contributions from transitions to
higher-lying states where final-state mixing is somewhat more
important. Experimentally, precise measurements of parity-odd
asymmetries are more practical for the total integrated gamma spectrum
rather than individual gamma transitions. But a calculation of the
asymmetry of the integral gamma spectrum requires an additional
averaging over the large number of distinct final states. In addition
the integral measurement also senses gamma cascades in addition to
single transitions. Parity-odd correlations in the integrated gamma
spectra of $^{35}$Cl, $^{81}$Br, $^{113}$Cd, $^{117}$Sn, and
$^{139}$La have previously been calculated by Flambaum and
Sushkov~\cite{pp:FLSU} and by Bunakov et. al.~\cite{pp:BUNAKOV}. However, more experimental
information on parity-odd asymmetries in integral gamma spectra from
heavy nuclei are needed in any attempt to make progress in this area.

We have searched for parity-odd directional \Gr~asymmetries in the
capture of cold polarized neutrons on \Al~, Cu, and \In~at the Los
Alamos Neutron Science Center (LANSCE). We have performed a simple
statistical estimate of the mean square value for the parity odd
asymmetries in these nuclei and obtain expected upper bounds which are
consistent with experiment. In addition, we performed measurements of
the directional \Gr~asymmetry for polarized cold neutron capture on
${\rm^{35}Cl}$ and on ${\rm ^{10}B}$. ${\rm^{35}Cl}$ is known to
possess a large parity-odd gamma asymmetry~\cite{pp:AVENIER,pp:VESNA}
and it is used to verify the sensitivity of our apparatus. ${\rm
^{10}B}$ is used extensively throughout the experiment, for neutron
shielding. Searches for parity-odd gamma asymmetries on several other
nuclei are in progress.

These measurements are being conducted in preparation for an
experiment to search for the parity violating \Gr~asymmetry in the
capture of polarized neutrons on protons by the NPDGamma
collaboration. The apparatus constructed for this measurement is
capable of measuring \Gr~ asymmetries with an accuracy of $10^{-8}$.

The remainder of the paper is organized as follows. We first provide a
brief theory section in which we outline the calculation and estimate
the expected root mean square of the \Gr~asymmetry in a current mode
\Gr~detector from the nuclei used in the experiment. We give a short
overview of the experimental layout and then describe the
measurements.  We conclude with a discussion about the results and the
associated implications.

\section{Theory and Statistical Estimates}

The simplest nuclear reaction which can produce a parity-odd
directional distribution of \Grs~is the capture of polarized neutrons
on protons. The differential cross-section in this simple system can
be calculated explicitly from the transition amplitudes of the
electro-magnetic part of the Hamiltonian between initial (capture) and
final (bound) two nucleon states, which possess mixed parity due to
the NN weak interaction.  In the \npdg\ reaction the primary process
is the strong interaction induced parity conserving {\em M1}
transition between the singlet and triplet S-wave states: $^{1}S_{0}$
, $^{3}S_{1}$. The weak interaction introduces a small parity
non-conserving admixture of P-wave states in the initial singlet
and the final triplet S-wave states. The largest contribution to the 
hadronic weak interaction comes from pion exchange and the measurement
of the parity-violating up-down asymmetry, $A_{\gamma}$, in the 
angular distribution of 2.2 MeV \Grs~with respect to the neutron 
spin direction (Eq.~\ref{eqn:CRSS}) , almost completely isolates 
the term proportional to the weak pion-nucleon coupling constant 
$f_{\pi}$~\cite{pp:ADLB}. 

\begin{equation} 
\frac{d\sigma}{d\Omega}\propto
\frac{1}{4\pi} \left(1+A_{\gamma}\cos\theta\right)\label{eqn:CRSS}
\end{equation} 
Here $\cos\theta$ is angle between the neutron spin direction 
and the \Gr\ momentum.

For the \npdg\ reaction, it can be shown that there is a simple
expression for the \Gr\ asymmetry in terms of the matrix elements
between initial and final states

\begin{equation}
A_{\gamma} \propto Re\frac{\epsilon\bra{^{3}P_{1}}{\bf E1}\ket{^{3}S_{1}}}
{\bra{^3S_1}{\bf M1}\ket{^1S_0}}~.
\end{equation}
Here 
\begin{equation}
 \epsilon = \frac{\bra{\psi_{\alpha'}}W\ket{\psi_{\alpha}}}{\Delta E}
\end{equation}
and $\alpha = \lbrace J,L,S,p \rbrace$ $(p = {\rm parity})$.

In heavy nuclei the interference term which produces the asymmetry is
much more complicated, involving many states.  Here, a neutron may
capture into an S or P wave state close to the neutron separation
energy $(S_n)$ and the weak interaction mixes the corresponding
amplitudes perturbatively.  For almost all nuclei except in few body systems
it is essentially impossible to calculate the parity violating
asymmetry from the strong and weak Hamiltonian, because of the large
number of \Gr\ transitions. However, because of the large number of
possible electromagnetic transitions in the compound nucleus the
calculation of the mean square matrix elements for the transition
amplitudes amounts to a summation of a large number of uncorrelated
random amplitudes which are approximately independent of the
transition energy for ${\mathrm E \leq S_n}$. One can then hope to use
statistical arguments to estimate the RMS value of the asymmetry from
nuclei close to a certain neutron separation energy.

Due to the large density of states close to the neutron separation
energy and the correspondingly small level spacing ${\mathrm D \simeq
\Delta E_c}$, parity violation is expected to be dominated by the
mixing of the two closest S and P wave states near $S_n$, in the
initial or capture state, and it is expected on general grounds that
parity violation due to mixing with lower lying states may be
neglected. The parity violating asymmetry comes from interference
between E1 and M1 gamma transitions. The \Gr~ asymmetry from the
decaying compound nucleus as measured in a current-mode gamma detector
is given by

%\begin{equation}
%A_{\gamma} = \xi\epsilon\frac{2\mathrm{Re}\sum_{J_f} \bra{J^{p}_f}{\bf H_{EM}}\ket{J^{p'}_i} \bra{J^{p}_i}{\bf H_{EM}}\ket{J^{p}_f}E^{4}_{\gamma,if}}
%{\sum_{J_f}\vert\bra{J^{p}_f}{\bf H_{EM}}\ket{J^{p'}_i}\vert^2 E^{4}_{\gamma,if}}~.\label{eqn:COMPASY}
%\end{equation}
\begin{equation}
A_{\gamma} = \epsilon\cdot\xi\cdot F(J_T,J_i)\frac{2\mathrm{Re}\left[\sum_{J_f} \bra{J^{p}_f}{\bf E1}\ket{J^{p'}_i} \bra{J^{p}_i}{\bf M1}\ket{J^{p}_f}E^{4}_{\gamma,if}\right]}
{\sum_{J_f}\left(\vert\bra{J^{p}_f}{\bf M1}\ket{J^{p}_i}\vert^2 + \vert\bra{J^{p}_f}{\bf E1}\ket{J^{p'}_i}\vert^2\right) E^{4}_{\gamma,if}}~.\label{eqn:COMPASY}
\end{equation}
Here the transitions are between initial ({\em i}) and final ({\em f})
compound nuclear states with total angular momentum $(J_{i},J_{f})$
and parity ({\em p},{\em p'}). $F(J_T,J_i)$ is the angular momentum coupling
factor resulting from the compound state polarization \cite{pp:FLSU}:
\[F(J_T,J_i) = (-1)^{2J_i+1/2+J_T}3(2J_i+1) \left\{ \begin{array}{ccc}
1 & 1/2 & 1/2 \\ J_T & J_i & J_i \end{array} \right\} \] 
$J_T$ is the angular momentum of the target nucleus before neutron
capture.

The dependence on the \Gr~ transition energy $E_{\gamma}$ in
eqn.~(\ref{eqn:COMPASY}) comes from the phase-space factor
$(E^{3/2}_{\gamma})$ in the transition amplitude and the linearity
$(\propto E_{\gamma})$ of the detector response as a function of
energy in a current mode gamma detector.  The factor $$\xi =
\frac{\sum_{f}I_{\gamma,if}E_{\gamma,if}}{S_n} \Rightarrow
\frac{1}{S_n}\frac{\int^{S_n}_{0}E^4_{\gamma}\rho_f(E_{\gamma})dE_{\gamma}}
{\int^{S_n}_{0}E^3_{\gamma}\rho_f(E_{\gamma})dE_{\gamma}}$$ arises
because the current mode gamma detector possesses no energy resolution
and therefore sees a superposition of currents from all transitions.
This has the effect of diluting the asymmetry $(0 < \xi \leq 1)$.
Here,

$$
I_{\gamma,if} = \frac{(\vert\bra{J^{p}_f}{\bf M1}\ket{J^{p}_i}\vert^2 + 
\vert\bra{J^{p}_f}{\bf E1}\ket{J^{p'}_i}\vert^2) E^3_{\gamma,if}}{
\sum_{J_f}(\vert\bra{J^{p}_f}{\bf M1}\ket{J^{p}_i}\vert^2 + 
\vert\bra{J^{p}_f}{\bf E1}\ket{J^{p'}_i}\vert^2)E^3_{\gamma,if}}
$$
is the relative intensity of a given transition.

\begin{figure}[h] 
  \includegraphics[scale=0.5]{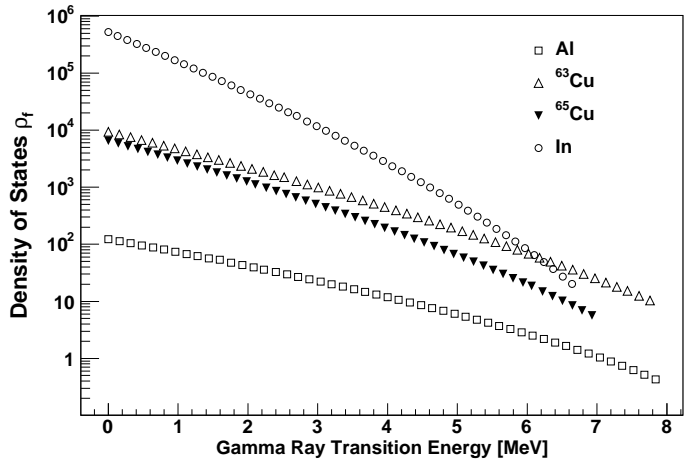}
  \caption{Density of final states in the excited compound nuclei
investigated in this work as a function of \Gr~ transition energy up
to the neutron separation energy. There are many more states at low
gamma energies than at high energies, and the decaying nucleus emits
many low energy gamma-rays before reaching the ground state. The level
density is calculated according to the Back-Shifted Fermi Gas Model.}
\label{fig:ALDST}
 \end{figure}

We estimate the density of final states using the Back-Shifted Fermi Gas
model (BSFGM) as~\cite{pp:EGBU,pp:DILG},
\begin{equation}
\rho_f(E_x) = \sum_{J}\frac{2J+1}{24\sqrt{2}\sigma^3 a^{1/4}}\frac{\exp[2\sqrt{a(E_x-\Delta)}-J(J+1)/2\sigma^2]}{(E_x-\Delta+t)^{5/4}}~,
\label{eqn:DEN}
\end{equation}
where $J$ is summed over $J_f-1,J_f,J_f+1$ for each final compound
nuclear state. Here, $a~{\rm [MeV^{-1}]}$ and $\Delta~{\rm [MeV]}$ are
determined from experimental data and the temperature parameter $t$ is
defined by $E_x-\Delta = at^2-t$.  $\sigma^2 = I_{eff}t/\hbar^2 \simeq
0.015 A^{5/3}t$ is the spin cut-off parameter and the {\em effective}
moment of inertia $I_{eff}$ takes on values between 50\% and 100\% of
the rigid body moment of inertia $I_{rig} = \frac{2}{5}MR^2$.  The
level density (Eq.~\ref{eqn:DEN}) is derived assuming random coupling
of angular momenta and the spin cut-off parameter arises as a result
of this treatment~\cite{pp:DILG}.  The excitation energy $E_x = S_n -
E_{\gamma}$ is the energy of the nucleus after the \Gr~transition from
the capture state. $E_x$ may be zero if the transition is to the
ground state.  Figure~\ref{fig:ALDST} shows the predicted density of
final states, using this model, for the \Al~, Cu, and \In, nuclei as a
function of \Gr\ energy.

For comparison and to estimate the uncertainty in the calculated 
asymmetries due to the model we also determine the asymmetries
using a constant temperature model (CTM) of the final density of
states~\cite{pp:EGBU}
\begin{equation}
\rho_f(E_x) = \frac{1}{T}\exp{(E_x-\Delta)/T}~.
\label{eqn:DENCT}
\end{equation}

The aim of this calculation is to find a simple ``generic'' formula
that holds for many nuclei and provides a good estimate of the size of
an asymmetry one can expect in a measurement of this nature.  The
denominator in eq.~\ref{eqn:COMPASY} is the parity allowed transition
from the initial compound state $(J_i)$ after capture of an S-wave
neutron. This transition has the largest amplitude and basically
determines the intensity of the gamma signal. In general {\em E1}
transitions outnumber {\em M1} transitions and the denominator is
primarily {\em E1} for most nuclei. We point out though, that if one
has initial states such that all or most parity allowed transitions
are {\em M1} in the range of expected \Gr~energies, as determined by
the density of states (as is the case for Al and In), then the
denominator would be {\em M1}. 

The root mean square of the \Gr~ asymmetry can then be estimated as
follows: We use the electric and magnetic dipole transition rates
which are given by 
\begin{eqnarray}
  &&\Gamma_{E1} = 2\pi \left<\left| \bra{J^{p'}_f}{\bf E1}\ket{J^{p}_i}\right|^2\right> \rho_f(S_n)  \nonumber \\
  &&\Gamma_{M1} = 2\pi \left<\left| \bra{J^{p}_f}{\bf M1}\ket{J^{p}_i}\right|^2\right> \rho_f(S_n) \label{eqn:TRNSAMPS}
\end{eqnarray}
respectively. The transition rates are strength functions. As the
density of states increases, the average matrix element squared
decreases and the transition rates are constant or slowly varying
functions of energy.

The root mean square of the detected intensity of the gammas that
depopulate the initial state is given by taking the average of the
squared denominator in eqn.~\ref{eqn:COMPASY}. Then, invoking the
randomness in the transition amplitudes (under the assumption that the
correlation is zero, so that the cross terms vanish), we find
\begin{eqnarray}
  && \left< \left( \sum_{J_f} \left(\left| \bra{J^{p'}_f}{\bf E1}\ket{J^{p}_i} \right|^2 + \left| \bra{J^{p}_f}{\bf M1}\ket{J^{p}_i} \right|^2\right) E^{4}_{\gamma,if}\right)^2\right> \nonumber \\ \nonumber
  &\simeq& \left(\sum_{J_f} \left<\left| \bra{J^{p'}_f}{\bf E1}\ket{J^{p}_i} \right|^2\right> E^{4}_{\gamma,if}\right)^2 + \left(\sum_{J_f} \left<\left| \bra{J^{p}_f}{\bf M1}\ket{J^{p}_i} \right|^2\right> E^{4}_{\gamma,if}\right)^2\nonumber \\ \nonumber
  &=& \left(\int^{S_n}_{0}E^{4}_{\gamma}\frac{\Gamma_{E1}}{2\pi \rho_f(S_n)}\rho_f(S_n)dE_{\gamma}\right)^2 + \left(\int^{S_n}_{0}E^{4}_{\gamma}\frac{\Gamma_{M1}}{2\pi \rho_f(S_n)}\rho_f(E_{\gamma})dE_{\gamma}\right)^2\nonumber\\ 
  &=& \frac{\Gamma^2_{E1}+\Gamma^2_{M1}}{4\pi^2 \rho^2_f(S_n)}\left(\int^{S_n}_{0}E^{4}_{\gamma}\rho_f(E_{\gamma})dE_{\gamma}\right)^2 \label{eqn:DENOM}
\end{eqnarray}
The factor $\rho_f(E_{\gamma})dE_{\gamma}$ arises in the
standard fashion, when converting the sum over final states into an
integral.

The root mean square of the interference term in the numerator
gives
\begin{eqnarray}
  & & 4\left<\left(\sum_{J_f} \bra{J^{p'}_f}{\bf E1}\ket{J^{p}_i} \bra{J^{p}_i}{\bf M1}\ket{J^{p}_f}E^{4}_{\gamma,if}\right)^2\right> \nonumber \\ \nonumber
  &\simeq& 4\sum_{J_f} \left<\left|\bra{J^{p'}_f}{\bf E1}\ket{J^{p}_i} \right|^2\right>\left<\left|\bra{J^{p}_i}{\bf M1}\ket{J^{p}_f}\right|^2\right> E^{8}_{\gamma,if} \nonumber \\ \nonumber
  &=&  4\int^{S_n}_{0}E^{8}_{\gamma}\frac{\Gamma_{E1}}{2\pi \rho_f(S_n)}\frac{\Gamma_{M1}}{2\pi \rho_f(S_n)}\rho_f(E_{\gamma})dE_{\gamma} \nonumber\\
  &=& \frac{\Gamma_{E1}\Gamma_{M1}}{\pi^2\rho^2_f(S_n)}\int^{S_n}_{0}E^{8}_{\gamma}\rho_f(E_{\gamma})dE_{\gamma}~. \label{eqn:NUMINT}
\end{eqnarray}
Where we again used the randomness in the transition amplitudes in 
going to the second line. With this, the mean square asymmetry can be 
estimated for each target from.
\begin{equation}
  {\sqrt{\expec{A^2_{\gamma}}}} \simeq 2F(J_T,J_i)\epsilon\xi\sqrt{\frac{\Gamma_{E1}\Gamma_{M1}}{\Gamma^2_{E1}+\Gamma^2_{M1}}\frac{\int^{S_n}_{0}E^{8}_{\gamma}\rho_f(E_{\gamma})dE_{\gamma}}{\left(\int^{S_n}_{0}E^{4}_{\gamma}\rho_f(E_{\gamma})dE_{\gamma}\right)^2}}~.
\label{eqn:MSASY}
\end{equation}
%\begin{equation}
%  {\expec{A}}^2_{\gamma} \simeq \frac{144\epsilon^2\xi^2}{S^{12}_n}\frac{Z(J_i,\frac{1}{2},\frac{1}{2},1,I)}{3\sqrt{2J_i+1}}\frac{\Gamma_{M1}}{\Gamma_{E1}}\int^{S_n}_{0}\frac{E^{10}_{\gamma}}{\rho(E_{\gamma})}dE_{\gamma}~.
%\label{eqn:MSASY}
%\end{equation}

To calculate the root mean square asymmetry for a particular nucleus,
one must then determine whether the transitions to the ground state
are mostly $E1$ or $M1$ and omit the corresponding amplitude in the
denominator. In the case of \Al~and \In~we then have
$\frac{\Gamma_{M1}}{\Gamma_{E1}}$, while for Cu we have
$\frac{\Gamma_{E1}}{\Gamma_{M1}}$.

Substituting the experimental value of the hadronic weak mean square
matrix element ($\Gamma_W = 1.8^{+0.4}_{-0.3}\times 10^{-7}~{\rm
eV}$)~\cite{pp:TRIPLE} and using $$\epsilon^2 =
\frac{\Gamma_{W}}{2\pi\rho_i}\frac{1}{D^2} \simeq \frac{\Gamma_W}{2\pi
D}$$ together with the fact that $E1$ transitions are approximately 10
times faster than $M1$ transitions, $\Gamma_{E1} \simeq 10
\Gamma_{M1}$~\cite{pp:KOPECK,pp:MCCULL}, the root mean square asymmetry can be
calculated for different nuclei and neutron separation energies.  When
evaluating eqn.~\ref{eqn:MSASY} for aluminum, for example, the single
particle level spacing is approximately $D \simeq 120 {\rm ~keV}$, the
dilution factor $\xi^2 \simeq 0.6$ and the ratio of integrals in
eqn.~\ref{eqn:MSASY} can be numerically evaluated to give $\simeq
6.5\times 10^{-2}$. The expected RMS value of the gamma asymmetry is then
about $1.3\times 10^{-7}$. The RMS \Gr\ asymmetry values and other
associated variables for the nuclei studied in this work are listed in
table~\ref{tbl:CALCRES}.

\begin{table}[hb]
\begin{center}
  \begin{tabular}{l|c|c|c|c|c|c|c|c|c} 
    \multicolumn{10}{c}{\bf Calculated RMS \Gr\ Asymmetries Values (BSFGM)} \\ \hline
                 & $S_n~\rm{[MeV]}$ &$J_T$&$J_i$& $F(J_T,J_i)$&$D~\rm{[eV]}$    &$\xi^2$ & $\epsilon^2$          & ${\rm I}$           &  $\sqrt{\expec{A^2_{\gamma}}}$\\ \hline
  ${\rm ^{27}Al}$&  8.0             &$5/2$&$2,3$  & 0.3       &$1.2 \times 10^5$&0.6     & $2.4 \times 10^{-13}$ & $6.5 \times 10^{-2}$&  $1.3 \times 10^{-7}$\\   
  ${\rm ^{63}Cu}$&  7.9             &$3/2$&$1,2$  & -0.4      &$4.8 \times 10^3$&0.5     & $6.0 \times 10^{-12}$ & $1.6 \times 10^{-3}$&  $1.4 \times 10^{-8}$\\   
  ${\rm ^{65}Cu}$&  7.1             &$3/2$&$1,2$  & -0.4      &$8.0 \times 10^3$&0.5     & $3.6 \times 10^{-12}$ & $2.7 \times 10^{-3}$&  $1.4 \times 10^{-8}$\\   
  ${\rm^{115}In}$&  6.8             &$9/2$&$4,5$  & 0.4       &400              &0.4     & $7.2 \times 10^{-11}$ & $7.3 \times 10^{-5}$&  $7.5 \times 10^{-8}$\\   
  \end{tabular} 													 
  \begin{tabular}{l|c|c|c|c|c|c|c|c|c} 
    \multicolumn{10}{c}{\bf Calculated RMS \Gr\ Asymmetries Values (CTM)} \\ \hline
                 & $S_n~\rm{[MeV]}$ &$J_T$&$J_i$& $F(J_T,J_i)$&$D~\rm{[eV]}$    &$\xi^2$ & $\epsilon^2$          & ${\rm I}$           &  $\sqrt{\expec{A^2_{\gamma}}}$\\ \hline
  ${\rm ^{27}Al}$&  8.0             &$5/2$&$2,3$& 0.3         &$1.2 \times 10^5$&0.7     & $2.4 \times 10^{-13}$ & $5.0 \times 10^{-2}$&  $1.3 \times 10^{-7}$\\   
  ${\rm ^{63}Cu}$&  7.9             &$3/2$&$1,2$& -0.4        &$4.8 \times 10^3$&0.6     & $6.0 \times 10^{-12}$ & $2.5 \times 10^{-3}$&  $1.8 \times 10^{-8}$\\   
  ${\rm ^{65}Cu}$&  7.1             &$3/2$&$1,2$& -0.4        &$8.0 \times 10^3$&0.6     & $3.6 \times 10^{-12}$ & $3.7 \times 10^{-3}$&  $1.7 \times 10^{-8}$\\   
  ${\rm^{115}In}$&  6.8             &$9/2$&$4,5$& 0.4         &400              &0.4     & $7.2 \times 10^{-11}$ & $9.9 \times 10^{-5}$&  $8.8 \times 10^{-8}$\\   
  \end{tabular} 													 
\end{center}
  \caption{RMS \Gr\ asymmetry values and associated variables, as
           estimated from the statistical approach $\left(I \equiv
           \int^{S_n}_0 dE_{\gamma} E^{8}_{\gamma}\rho(E_{\gamma})/\left(\int^{S_n}_0 dE_{\gamma} E^{4}_{\gamma}\rho(E_{\gamma})\right)^2 {\rm ,}~D\equiv D_o\sum 2J_i + 1 \right)$, $D_o$ was taken from~\cite{bb:MUGH1,bb:MUGH2}.}\label{tbl:CALCRES}
\end{table}

\subsection{Theory Discussion}

The results in tables~\ref{tbl:CALCRES} and~\ref{tbl:FNASYVALS} show
no large enhancements. There are several reasons why one may expect this
behavior. For example, the levels are highly degenerate, the sign
of the asymmetry is random and the transitions mix incoherently,
producing a $1/\sqrt{N}$ suppression. There is also no $kR$
enhancement for the direct capture calculations done here, which are
appropriate for the low neutron energies used in these experiments.

In~\cite{pp:FLSU} Flambaum and Sushkov calculated the average value
for the integral \Gr~spectrum relative to the S-wave amplitude for
thermal neutrons $$a_o =
\frac{g}{4k^2}\frac{T^2_s\Gamma^{(\gamma)}_{eff}}{(E-E_s)^2+\frac{1}{4}\Gamma^2_s}$$
which is far from p-wave resonance so that it's contribution to the
cross section can be neglected. The root mean square asymmetry is
given, in their notation, by

\begin{equation}
  \expec{A_9} = -2Re\left( \frac{\epsilon}{E-E_p-\frac{1}{2}i\Gamma_p}\right)\frac{F(J_T,J_i)}{3\sqrt{2J_i+1}}r~. \label{eqn:FLAMASY}
\end{equation}
Where $(r)$ is an integral over the {\em E1} and {\em M1} radiative
strength functions, detection efficiency and density of final states,
corresponding to our integral in eqn.~\ref{eqn:NUMINT}. Equation~\ref{eqn:FLAMASY} 
may be compared to our result above. Flambaum and
Sushkov also state that the~\Gr~asymmetry arises as a result of the
{\em E1,M1} interference, that the transitions are random, and that
the asymmetry is statistically suppressed after averaging.

The main difference between our calculations and those done by
Flambaum and Sushkov is that they consider a p-wave resonance near the
thermal (or cold) region mixed by parity violation with one S-wave
resonance, while our treatment takes account of all S and P-wave
resonances, but in the tail, far from resonance, at the average
spacing D or more.

\section{Experiment}

The NPDGamma apparatus used for the measurements is located on flight
path 12 at the Manuel Lujan Jr. Neutron Scattering Center at
LANSCE. The LANSCE linear accelerator delivers $800$~MeV protons to a
storage ring, which compresses the beam to $250$~ns wide pulses at the
base. The protons from the storage ring are incident on a split
tungsten target at a rate of $20$~Hz and the resulting spallation
neutrons are cooled by and backscattered from a cold ${\rm H_2}$
moderator with a surface area of $12 \times 12$~${\rm cm^2}$ .  For
the measurements described here, the cold neutrons were transported to
the experimental apparatus by a neutron guide and then transversely
polarized by transmission through a polarized $^{3}$He cell. Three
$^{3}$He ion chambers were used to monitor beam intensity and
polarization.  A radio frequency spin flipper was used to reverse the
neutron spin direction on a pulse by pulse basis. The polarized
neutrons then captured on a target placed in the center of the gamma
detector array.  The gamma rays from the neutron capture were detected
by an array of 48 CsI(Tl) detectors operated in current
mode~\cite{pp:DET,pp:swb}. The entire apparatus was in a homogeneous
10 Gauss field, which was required to maintain the neutron spin
downstream of the polarizer, with a gradient of less than $1$~mG/cm to
make spin-dependent Stern-Gerlach steering of the polarized neutron
beam negligible.

\begin{figure}[h]
  %\hspace{2cm}angle = 315,
  \includegraphics[scale=0.4]{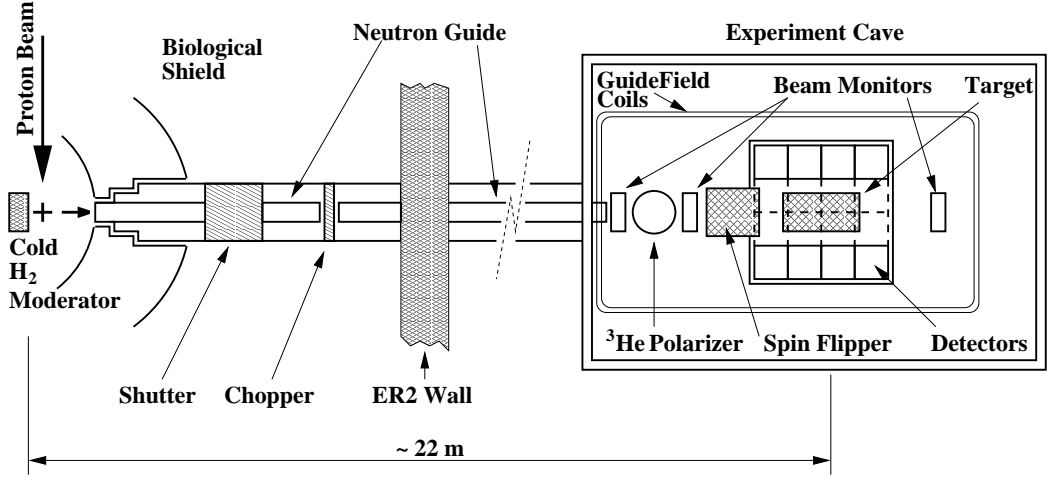}
  \caption{Schematic of the experimental setup.}
  \label{fig:EXPSTP} 
\end{figure} 

\begin{figure}[h]
  \includegraphics[scale=0.4]{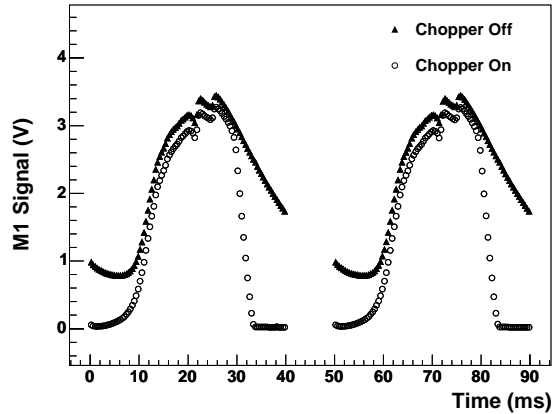}
  \caption{Normalized signal from the first beam monitor downstream
           of the guide exit. The solid triangles show the signal
           obtained from a run where the chopper was parked open. The
           open circles corresponds to a run taken with the chopper
           running.} \label{fig:MON1SIG}
\end{figure}

Figure~\ref{fig:EXPSTP} shows the flight path and experimental setup.
The distance between the moderator and target is about 22 meters.  The
flight path 12 beam line consists of a neutron guide, a shutter, and a
beam chopper. The pulsed spallation neutron source allowed us to know
the neutron time of flight or energy accurately.  The chopper is used
to define the time of flight frame and to prevent neutrons from
different frames to mix and thus dilute the neutron energy
information. In this experiment the chopper was used to close the beam
line before the end of the frame which allowed us to take beam-off
(pedestal) data for $\simeq 6$~ms at the end of each neutron pulse
which is needed for detector pedestal and background studies
(Fig.~\ref{fig:MON1SIG}). The last $10$~ms after sampling stops is
used by the DAQ for data transfer.  A detailed description of the FP12
neutron guide and performance is given in~\cite{pp:pil}. The measured
moderator brightness has a maximum of $1.25 \times 10^8$ ${\rm
n/(s\cdot cm^{2}\cdot sr\cdot meV\cdot \mu A)}$ for neutrons with an energy of $3.3$~meV.

The neutrons were polarized by passing through a 12 cm diameter glass
cell containing polarized ${\rm^3He}$ (~\cite{pp:chupp,pp:pol} and references
within).  The beam polarization was measured with the beam monitors
using neutron transmission (${\rm^3He}$ polarization can be monitored
using NMR). For gamma asymmetry measurements, the figure of merit is
the statistical accuracy that can be reached for a certain running
time, which is proportional to the product $P_n \sqrt{T_n}$, where
$T_n$ is the neutron transmission through the cell and $P_n$ is the
neutron polarization ~\cite{pp:Jones}.  The neutron transmission
increases with energy whereas the neutron polarization decreases with
energy. In the analysis of the data the neutron polarization was
calculated separately for each run by fitting the transmission
spectrum to the expression $P_{n} = \tanh{(\sigma_c n l P_{He})}$,
using a ${\rm ^3He}$ thickness of $4.84~{\rm bar\cdot cm}$, which was
separately measured. Here, $\sigma_c = \sigma_o/\sqrt{E}$ with neutron
energy $E$ in units of meV, and $\sigma_o = 27168~{\rm b},~nl =
4.84\cdot 2.688\times10^{23}~{\rm atoms/m^2}$.

%\begin{figure}[h]
%  \includegraphics[scale=0.35]{./NeutronPol.eps}
%  \includegraphics[scale=0.35]{./NeutronTrans.eps}
%  \caption{Left: Plot of beam polarization as a function of neutron
%           time of flight. Right: Neutron transmission data for an
%           unpolarized ${\rm ^3He}$ spin filter cell and a polarized
%           cell. The chopper cutoff is completed just below 35 ms
%           tof.} \label{fig:NPOL}
%\end{figure}

The primary technique for reducing false asymmetries generated by gain
non-uniformities, slow efficiency changes and beam fluctuations is
frequent neutron spin reversal. This allows asymmetry measurements to
be made in each spin state for opposing pairs of detectors and for
consecutive pulses with different spin states, thereby suppressing the
sensitivity of the measured asymmetry to detector gain differences,
drifts, and intensity fluctuations. By carefully choosing the sequence
of spin reversal, the linear and quadratic components of
time-dependent detector gain drifts in a sequence can be greatly
suppressed. To achieve the neutron spin reversal, the experiment
employed a radio frequency adiabatic neutron spin rotator
(RFSR)~\cite{pp:spfl} which operates at $29$-kHz for the 10~G guide
field. The neutron spin direction is reversed when the RFSR is on and
is unaffected when it is off.  The spin flip efficiency averaged over
the beam cross-section ($~5$~cm diameter) was measured to be about
99\%

The polarized neutrons then captured on a target placed in the center
of the gamma detector array. The targets were thick enough to stop
most of the neutron beam by capture or scattering with diameters
larger than the beam cross section.  The capture $\gamma$ signals from
all of the targets measured were large compared to noise and
background.

%The ${\rm CCl_4}$ target and the mount for the aluminum and
%copper target sheets are shown in Fig.~\ref{fig:TARGTS1}.
%\begin{figure}[h]
%  \includegraphics[scale=0.3]{./ccltarget_01.eps}
%  \hspace{3cm}
%  \includegraphics[bb=80 0 850 590,clip,scale=0.25]{./altarget_01.eps} %bb=0 0 800 400,clip,
%  \caption{Left: Container for the liquid ${\rm CCl_4}$ target
%           material.  The target material volume is $\sim 33~{\rm
%           cm^3}$. The container is made of Teflon and is closed at
%           the top with a Teflon lid and Teflon screws. Right: Al and
%           Cu target holder and target sheets. The thickness of the
%           target can be adjusted by adding or removing sheets.}
%  \label{fig:TARGTS1}
%\end{figure}

The housing for the $33~{\rm cm^3}$ liquid ${\rm CCl_4}$ target was
made of Teflon.  The ${\rm CCl_4}$ liquid is 99.9\% chemically pure,
with less than 0.01\% water content. The aluminum and copper targets
consisted of a number of sheets supported by an aluminum frame.  Each
target sheet is an approximately $1$~mm thick square with $8.5$~cm
sides. The arrangement of the target into sheets with a gap between
the sheets reduced $\gamma$ attenuation in the target. The total
length of the target (including gaps) was 30 cm. Target out background
runs and runs with the empty frame were conducted as well and the
background is taken into account in the final determination of the
asymmetry (see table~\ref{tbl:BCKGR}).  The boron target consists of a
1 cm thick 15 cm by 15 cm sheet of sintered \BC~ glued to an aluminum
holder consisting of a simple (thin) aluminum sheet. The indium target
was approximately 12 mm thick, covering a circular cross-sectional
area with a radius of $\sim 3$~cm at the center of the beam. For each
of the targets the beam was collimated to a diameter of about $5~{\rm  cm}$.

%The
%target positions along the beam, relative to the detector array, are
%shown in Fig~\ref{fig:DTGEO}. In the beam left and right directions
%the detector array and targets are centered on the beam guide axis.

%\begin{figure}[h]
%  %\hspace{1cm}
%  \includegraphics[scale=0.5]{./DetTargetGeoms2.eps} %bb=0 0 800 400,clip,
%  \caption{Target positions with respect to the detector array, along
%           the beam.}\label{fig:DTGEO}
%\end{figure}

\begin{table}
\begin{center}
\begin{tabular}{lrr}
\hline
  \multicolumn{3}{c}{{\bf Relative Background}} \\
\hline 
\BC   & $~~~~~\leq$&$17\%$ \\
Al    & $~~~~~\leq$&$15\%$ \\
In    & $~~~~~\leq$&$11\%$ \\
\CCL  & $~~~~~\leq$&$ 8\%$ \\
Cu    & $~~~~~\leq$&$ 7\%$ \\
\hline
\end{tabular}
\end{center}
\caption{Targets with their relative background contributions (target
         in versus target out). In each case the maximum value is
         stated for the detector with the largest background signal.
         The relative amount of background varies, because the magnitude 
	 of the \Gr~ signal varies with target while the target-out
         background remains constant.}
\label{tbl:BCKGR}
\end{table}

\begin{table}[hb]
  \begin{center}
    \begin{tabular}{llr} \hline
      & $\frac{2\sigma_{inc}}{3\sigma_{tot}}$  & $\langle \Delta_{dep}(t_i) \rangle$ \\
      \hline  
      Al   & $~~~3\times10^{-3}$	& 1    \\
      Cu   & $~~~2\times10^{-2}$	& 0.95 \\
      \CCL & $~~~7\times10^{-2}$	& 0.95 \\
      In   & $~~~2\times10^{-3}$   & 1    \\
      \BC  & $~~~5\times10^{-4}$   & 1    \\
      \hline
    \end{tabular}
    \end{center}
  \caption{Spin-flip probability estimate and corresponding
    corrections to the asymmetry due to depolarization.}
  \label{tbl:SFAC}
\end{table}

The depolarization of neutrons {\em via} spin flip scattering from
the nuclei dilutes the asymmetry. For all targets the neutron
depolarization is a small effect which can be estimated to sufficient
accuracy for nonmagnetic materials using the known neutron coherent
and incoherent cross sections.  Table~\ref{tbl:SFAC} lists the
estimated spin-flip probabilities for the targets used and the
corresponding calculated average correction factors $\langle F(t_i)
\rangle$.  The degree of spin flip scattering is neutron energy
dependent and a Monte Carlo calculation for the depolarization as a
function of neutron energy was applied to the data.

The detector array consists of 48 CsI(Tl) cubes arranged in a
cylindrical pattern in 4 rings of 12 detectors each around the target
area (Fig.~\ref{fig:det}).  In addition to the conditions set on the
detector array by the need to preserve statistical accuracy and
suppress systematic effects, the array was also designed to satisfy
criteria of sufficient spatial and angular resolution, high
efficiency, and large solid angle coverage~\cite{pp:DET}.  Because of
the possible small size of the asymmetries and the proposed
measurement accuracy the average rate of neutron capture and the
corresponding gamma rate in the detectors must be high to keep the
run-time reasonable. Because of the high rates and for a number of
other reasons discussed in~\cite{pp:DET}, the detector array uses
current mode gamma detection. Current mode detection is performed by
converting the scintillation light from CsI(Tl) detectors to current
signals using vacuum photo diodes (VPD), and the photocurrents are
converted to voltages and amplified by low-noise solid-state
electronics~\cite{pp:swb}.

%\begin{figure}[h]
%  \includegraphics[scale=0.35]{./cntstats.eps}
%  \caption{Counting statistics analysis results for a typical
%           detector module.  The RMS width from counting statistics is
%           compared to the width seen from pedestal runs (electronic
%           noise). A fit to the beam on, data histogram with target
%           shows an RMS width of $6.1~\pm 0.04$~mV.}
%           \label{fig:CNTSTATS}
%\end{figure}

In current mode detection, the counting statistics resolution is
limited by the RMS width in the sample distribution. For our detector
array this width is dominated by fluctuations in the number of
electrons produced at the photo-cathode of the VPD, which is dominated
by \Gr~counting statistics when the beam is on. During beam on
measurements, the shot noise RMS width is then given by~\cite{bb:DVR}
\begin{equation}
 \sigma_{I_{\mathrm{shot}}} = \sqrt{2qI}~\sqrt{f_B}, \label{eqn:SHN}
\end{equation} 
where $q$ is the amount of charge created by the photo cathode per
detected gamma-ray, $I$ is the average photo-current per detector and
$f_B$ is the sampling bandwidth, set by the $0.4$~ms time bin width in
the time of flight spectrum~\cite{pp:DET,pp:gericke2}.  

%Figure~\ref{fig:CNTSTATS} shows the RMS
%width for a typical detector, as seen at the preamplifier output. The
%width from counting statistics is compared to the RMS width seen for
%beam-off electronic noise.

\section{Analysis and Results}

\subsection{Asymmetry Definition}

\begin{figure}[h]
  \includegraphics[scale=0.5]{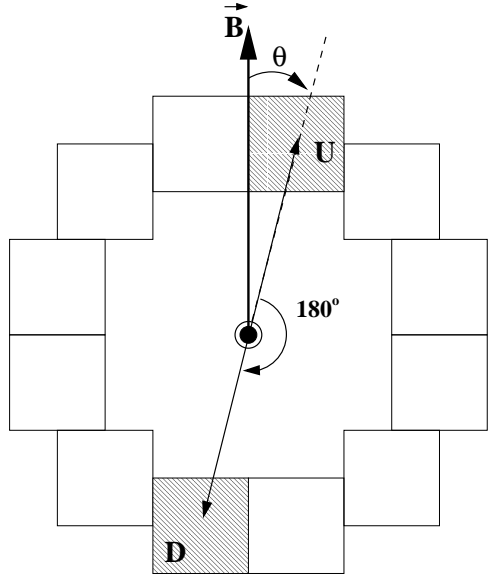}
  \caption{A ring of detectors and one up-down pair, as seen with beam
           direction into the page. $\vec{B}$ is the magnetic holding
           field defining the direction of the neutron polarization.}
  \label{fig:det}
\end{figure}

For a point target and a detector array with perfect spatial
resolution, the measured $\gamma$-ray angular distribution would be
proportional to the differential cross section $Y = 1 +
A_{\gamma}~\cos{\theta}$, where $\theta$ is the angle between the
neutron polarization and the momentum of the emitted photon and $
A_{\gamma,UD}$ is the parity-odd up-down (UD) asymmetry. A third term
is present if a parity-conserving (PC) left-right (LR) asymmetry
exists~\cite{pp:CGP}. In that case $Y = 1+A_{\gamma,UD}~\cos{\theta} +
A_{\gamma,LR} ~\sin{\theta}$. However, the relationship between the
basic expression for the \Gr~ yield and the measured asymmetry is
complicated by a number of small neutron energy dependent effects.  A
separate asymmetry is calculated for each detector pair, as defined in
Fig~\ref{fig:det}.

The physics asymmetry for a given detector pair (p), spin sequence
(j), and neutron time of flight $(t_i)$ is given by
\begin{eqnarray}
  \left(A^{j,p}_{UD}(t_i) + \beta A^{j,p}_{UD,b}(t_i)\right)\langle G_{UD}(t_i) \rangle &+& 
  \left(A^{j,p}_{LR}(t_i) + \beta A^{j,p}_{LR,b}(t_i)\right)\langle G_{LR}(t_i) \rangle \nonumber \\ 
  &=& \frac{\left(A^{j,p}_{raw}(t_i) - A^p_g A_f(t_i) - A^p_{noise}\right)}{P_n(t_i)\Delta_{dep}(t_i)\Delta_{sfl}(t_i)} \label{eqn:TBPHASY} \\ \nonumber
\end{eqnarray}

Here, $A^{j,p}_{raw}(t_i)$ is the measured asymmetry.  The background
asymmetries $(A^{j,p}_{UD,b}, A^{j,p}_{LR,b})$ and the relative signal
level $(\beta)$ must be measured in auxiliary measurements. $A^p_g$ is
the gain asymmetry between the detector pair and $A_f(t_i)$ is the
asymmetry from pulse to pulse beam fluctuations.  The neutron energy
and detection efficiency weighted spatial average detector cosine
(up-down asymmetry) with respect to the (vertical) neutron
polarization is given by $\langle G_{UD}(t_i) \rangle \simeq
cos(\theta)$, while the detector sine (left-right asymmetry) is given
by $\langle G_{LR}(t_i) \rangle \simeq sin(\theta)$. These
detector-target geometry corrections have been modeled for each target
geometry.  Also included are the correction factors due to the neutron
beam polarization $(P_n(t_i))$, the spin flip efficiency $(\Delta_{sfl}(t_i))$
and the neutron depolarization in the target $(\Delta_{dep}(t_i))$.

The measured ({\em raw}) asymmetry $(A^{j,p}_{raw})$ for each detector
pair and neutron energy can be extracted in the usual way, by forming
a ratio of differences between cross-sections to their sum.  However,
to suppress first and second order detector gain drifts~\cite{pp:bow}
the raw asymmetries were formed for all valid sequences of 8 macro
pulses with the correct neutron spin state pattern as shown in
eqn.~(\ref{eqn:ASY}).
\begin{equation}
   A^{j,p}_{raw}(t_i) = \frac{\sum_{s=\uparrow}(U_s(t_i)-D_s(t_i))-\sum_{s=\downarrow}(U_s(t_i)-D_s(t_i))}
{\sum_{s=\uparrow}(U_s(t_i)+D_s(t_i))+\sum_{s=\downarrow}(U_s(t_i)+D_s(t_i))}.
  \label{eqn:ASY}
\end{equation}
Here the sum is over all four signals with the corresponding spin
state in a spin sequence for the up (U) and down (D) detector in a pair. A
so-called valid 8 step sequence of spin states is defined as
($\uparrow\downarrow\downarrow\uparrow\downarrow\uparrow\uparrow\downarrow$).
Asymmetries were measured for 55 different neutron energies between
approximately 2 and 16 meV, with a resolution of $\sim 0.2~{\rm
to}~1.0~{\rm meV}$ per time bin, respectively.

It is important to realize that signal fluctuations that are not
correlated with the switching of the neutron polarization direction,
such as beam and detector gain fluctuations, will average out and
don't contribute to the asymmetry. It is, however, essential that
these signals have an RMS width that is small compared to the RMS
width in the asymmetries of interest (driven by counting statistics)
so that they do not reduce the statistical significance of the result
and are averaged to zero quickly compared to the time it takes to
measure the asymmetry to the desired accuracy. Possible false
asymmetries due to spin-state correlated electronic pickup (additive
asymmetry) and possible magnetic field induced gain changes
(multiplicative asymmetry) in the detector VPDs have previously been
measured and are consistent with zero to within $5 \times
10^{-9}$~\cite{pp:DET}.

The detector pair physics asymmetries as represented by
eqn.~\ref{eqn:TBPHASY} can then be combined in error weighted averages
over the neutron time of flight spectrum to form a single asymmetry
for each detector pair in the array, for a single 8-step sequence of
beam pulses.  If beam intensity levels are sufficiently stable over the
measurement time these sequence asymmetries can be histogrammed for
each pair.  Typical run lengths were $\sim 8.3$ minutes and included
10000 beam pulses or 1250 8-step sequences and the asymmetry
measurements performed usually extended over several hundred
runs.

\subsection{Results}

\begin{figure}[ht]
  \hspace{0cm}
  \includegraphics[scale=0.33]{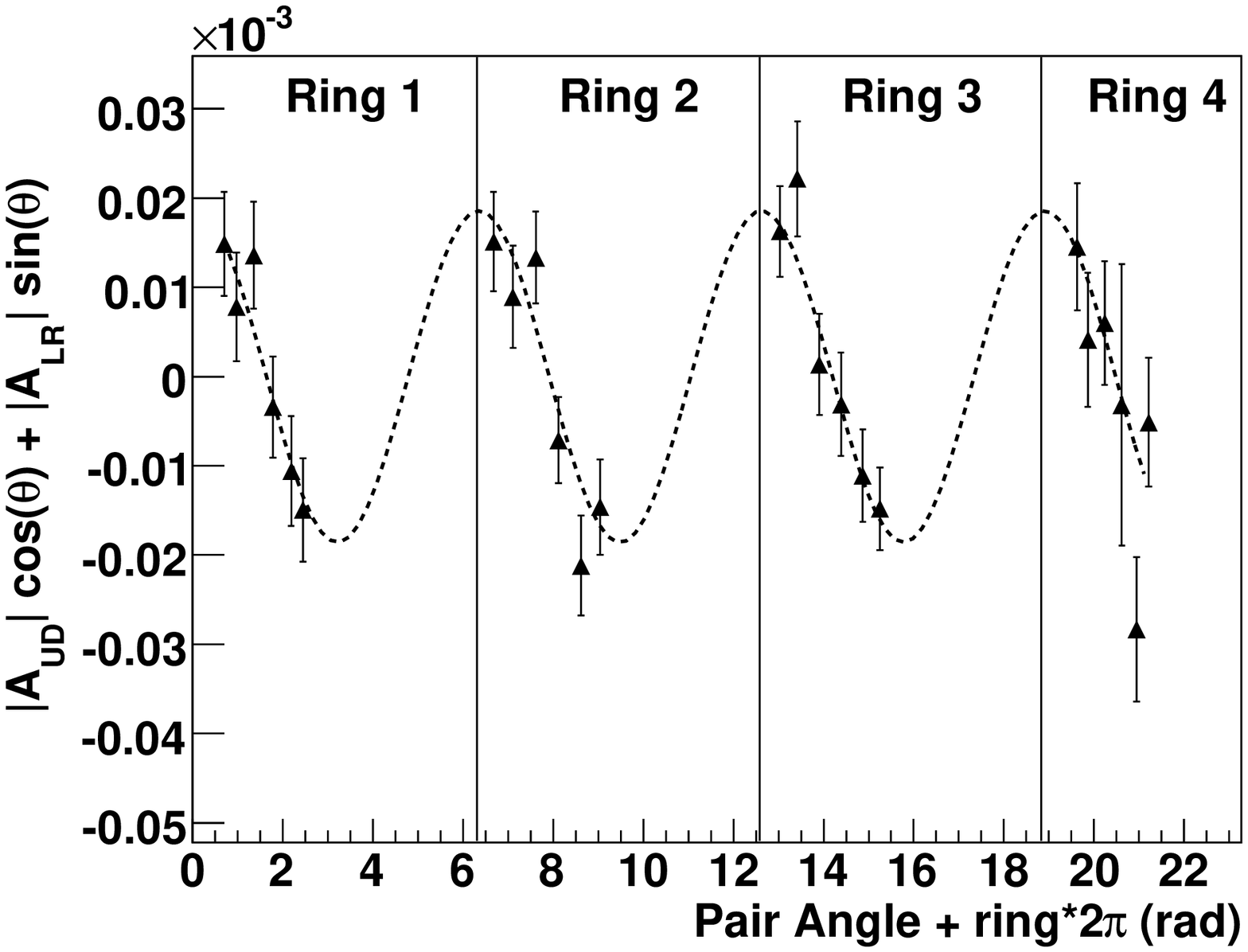}
  \includegraphics[scale=0.35]{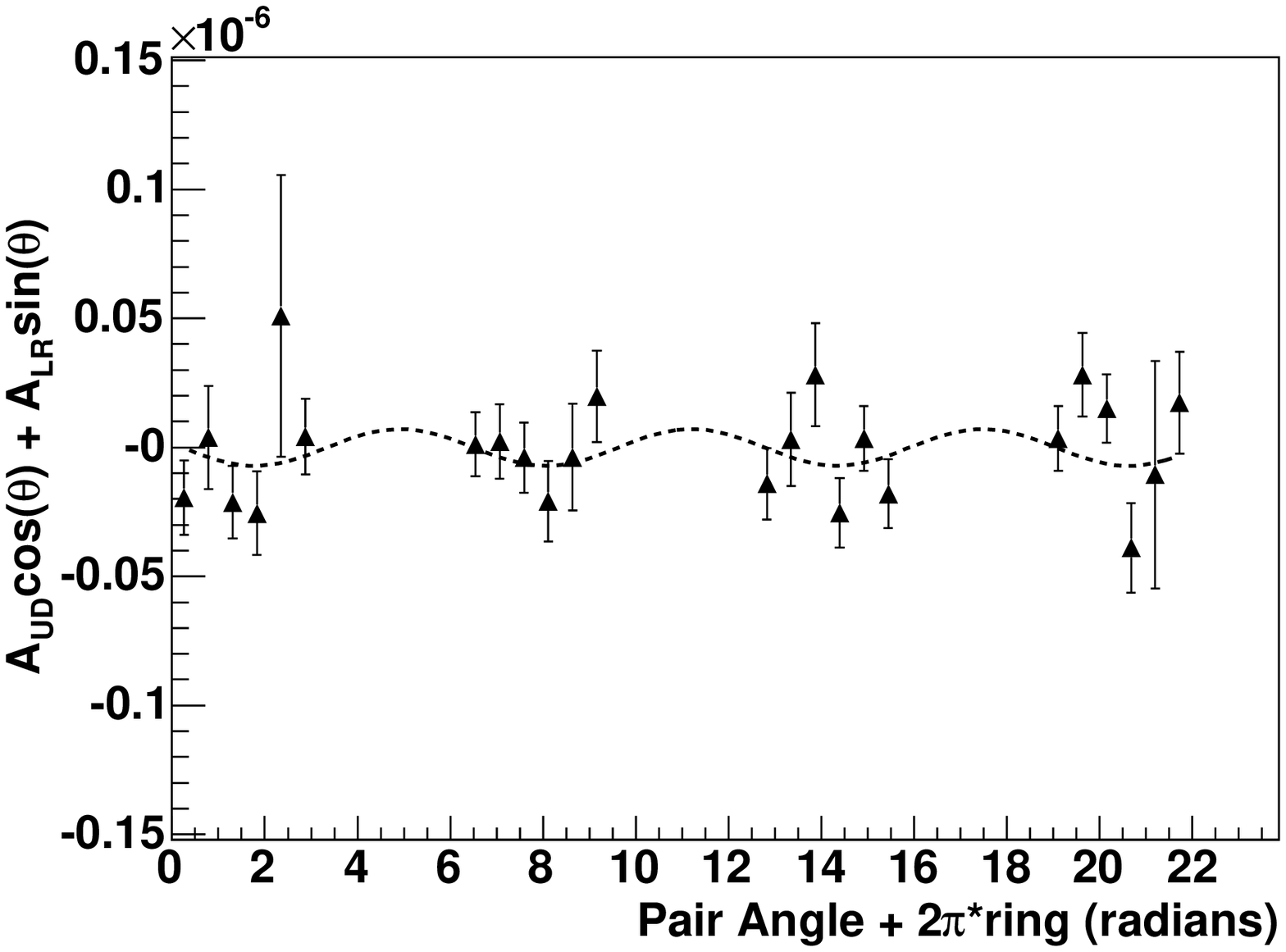}
  \caption{Left: \CCL~asymmetries for each pair, plotted versus angle
           of the first detector in the pair w.r.t the vertical. The
           total array asymmetry is extracted from the fit. Right:
           Noise asymmetries.}\label{fig:CCLPRASYS}
\end{figure}

The known parity-odd gamma asymmetry in \CCL~was used to verify that a
nonzero asymmetry can be measured with our apparatus.  The
\CCL~asymmetry was also used to verify the geometrical dependence of
the pair asymmetries. For this purpose all 24 pair asymmetries,
extracted from the histogrammed 8-step sequence asymmetries over all
data obtained with that target, were multiplied by their mean geometry
factors and plotted versus their corresponding mean angle.  The
resulting graph is shown in Fig.~\ref{fig:CCLPRASYS}. The fit function
used to extract the total array asymmetry is $A_{UD}\cos{\theta} +
A_{LR}\sin{\theta}$.

\begin{table}[hb]
\begin{center}
  \begin{tabular}{l|r|r|r} 
    \multicolumn{4}{c}{${\bf Asymmetries~and~RMS~width}$} \\ \hline
                 &{\bf Up-Down}                             &{\bf Left-Right}                         &{\bf RMS width}       \\
                 &                                          &                                         &{\bf (typ.)}          \\ \hline
    Al           & $\left(-0.02 \pm  3\right)\times10^{-7}$ & $\left(-2   \pm  3\right)\times10^{-7}$ & $1.2\times10^{-3}$   \\   
    \CCL         & $\left(-19   \pm  2\right)\times10^{-6}$ & $\left(-1   \pm  2\right)\times10^{-6}$ & $1.0\times10^{-3}$   \\  
    \BC          & $\left(-1    \pm  2\right)\times10^{-6}$ & $\left(-5   \pm  3\right)\times10^{-6}$ & $0.7\times10^{-3}$   \\ 
    Cu           & $\left(-1    \pm  3\right)\times10^{-6}$ & $\left( 0.3 \pm  3\right)\times10^{-6}$ & $1.0\times10^{-3}$   \\  
    In           & $\left(-3    \pm  2\right)\times10^{-6}$ & $\left( 3   \pm  3\right)\times10^{-6}$ & $0.4\times10^{-3}$   \\
    Noise (add.) & $\left( 2    \pm  5\right)\times10^{-9}$ & $\left(-7   \pm  5\right)\times10^{-9}$ & $2.0\times10^{-6}$   \\
    Noise (mult.)& $\left( 3    \pm  7\right)\times10^{-9}$ & $\left(-9   \pm  7\right)\times10^{-9}$ & $0.2\times10^{-3}$   \\
    Beam$\cdot$Gain    & N/A                                      & N/A                                     & $1.0\times10^{-5}$   \\
  \end{tabular} 
\end{center}
  \caption{Up-Down and Left-Right asymmetries for the target
           materials. Stated errors are statistical only. The RMS
           widths are taken from histograms with single 8-step
           sequence asymmetries for a detector pair as individual
           entries.}\label{tbl:FNASYVALS}
\end{table}

In general, the up-down and left-right asymmetries must be extracted
using the fit described above. Higher order corrections to the fitting
function used here (parity violating or not) are introduced by higher
partial waves in the expansion of the initial and final two nucleon
states representing more complicated scalar combinations between the
neutron spin ${\bf s}_n$ and outgoing \Gr\ momentum direction ${\bf
k}_{\gamma}$.  For the up-down asymmetry the angular distribution is
obtained from initial and final two-nucleon states with components up
to the P-waves producing the $ {\bf s}_n \cdot {\bf k}_{\gamma} $
correlation. The left-right asymmetry originates from the $ {\bf
s}_n\cdot({\bf k}_{\gamma}\times{\bf k}_n)$ correlation. Parity
violating corrections from higher partial waves are negligible because
they represent a second order perturbation proportional to the weak
coupling squared. The results of the asymmetry measurements are
summarized in table~\ref{tbl:FNASYVALS}. Note that beam asymmetries
are only produced if there are pulse to pulse fluctuations in the
number of neutrons and only in combination with a difference in gain
between a given detector pair. Neither beam fluctuations nor detector
gain differences are correlated with the neutron spin and therefore
the Beam$\cdot$Gain asymmetry does not contain any up-down or
left-right dependence. Due to the sum over the eight step sequence,
the Beam$\cdot$Gain asymmetry is zero and its root mean square width
is determined by the size of beam fluctuations. The additive and
multiplicative noise asymmetries in ~\ref{tbl:FNASYVALS} are measured
without a light signal from the detectors (electronic noise only) and
with a light signal from LEDs embedded in the detectors respectively.
The large RMS width for the multiplicative noise asymmetry is a result
of larger fluctuations with LEDs~\cite{pp:DET}.

\subsection{Errors}

The final statistical errors stated in table~\ref{tbl:FNASYVALS} are
taken from the distribution of sequence values $\sigma^2_{\gamma}/N =
(E(A^2_{\gamma}) - E(A_{\gamma})^2)/N$, with $N$ histogrammed 8-step
sequence asymmetries. Any non-random effect such as those introduced
by the correction factors $|\langle G(t_i), \rangle|,~
\Delta_{dep}(t_i),~P_n(t_i),~\Delta{sfl}(t_i)$ are treated as systematic errors. These
enter as
$$
\sigma_{\gamma, _{Sys}} = A_{\gamma}\sqrt{\left(\frac{\sigma_{P_n}}{P_n}\right)^2+
\left(\frac{\sigma_{sfl}}{\Delta_{sfl}}\right)^2 + \left(\frac{\sigma_{G}}{G}\right)^2 + 
\left(\frac{\sigma_{dep}}{\Delta_{dep}}\right)^2}
$$ 
and are added in quadrature with the statistical error.

The errors on the beam polarization and spin flip efficiency were
calculated to be 4\% and 10\% respectively.  The error on the geometry
factor is estimated to be less than 1\% from variations observed in
the values when varying the step size in the Monte Carlo, simulating
\Gr~ interaction in the detectors.  The error on the spin flip
scattering is estimated to be on the order of a few percent. Since the
systematic errors are scaled by the asymmetry, their contribution to
the overall error on the asymmetry is negligible compared to the
statistical error, except for the case of the \CCL~ target, which has
a large non-zero asymmetry.  For \CCL~, the systematic error is
$\simeq 2.3\times10^{-6}$.  So the \CCL~ Up-Down physics asymmetry and
its total error is $(-19 \pm 3)\times10^{-6}$. A previous measurement
of this asymmetry by this collaboration found $(-29.1 \pm
6.7)\times10^{-6}$ ~\cite{pp:greg}.  M. Avenier and
collaborators~\cite{pp:AVENIER} found an Up-Down asymmetry for ${\rm
^{35}Cl}$ of $(-21.2 \pm 1.7)\times10^{-6}$ , while V.A. Vesna and
collaborators found $(-27.8 \pm 4.9)\times10^{-6}$ ~\cite{pp:VESNA}
(see also~\cite{pp:KRUP}).

\section{Conclusion}

The NPDGamma collaboration has searched for \Gr~ asymmetries from
polarized slow neutron capture on \Al, Cu, \In~ and \BC. The asymmetry
measurements for these targets were consistent with zero at the few
$10^{-7}$ level for \Al~ and at the few $10^{-6}$ level for Cu and
\In~. All asymmetries are consistent with zero within errors. The
${\rm ^{35}Cl}$ asymmetries obtained from the \CCL~ measurements are
consistent with results from previous measurements. A statistical
model, in combination with previous measurements of weak matrix
elements in compound nuclei, was used to estimate the expected RMS size of the
parity violating \Gr~ asymmetries in \Al, Cu, and \In~. Based on this
model it is expected that non-zero measured asymmetries will be
smaller than the estimated width 68.3\% of the time.  The upper bounds
on the measured asymmetries are therefore consistent with the
estimates obtained from these statistical calculations.  Based on the
inverse relationship between the single particle level spacing and the
size of the asymmetry, one would expect a large number of very small
or essentially zero asymmetries when performing measurements for many
larger nuclei, but one would also expect to find a small number of
nuclei with enhanced asymmetries. We plan to continue measurements in
other nuclei in the mass range ${\rm A > 50}$ to test this hypotheses
more precisely and to further investigate the predictions of the
statistical approach to parity violation in compound nuclei.

\section{Acknowledgments}
The authors would like to thank Mr.\ G.\ Peralta (LANL) for his
technical support during this experiment, Mr.\ W.\ Fox (IUCF) and
Mr.\ T.\ Ries (TRIUMF) for the mechanical design of the array and the
construction of the stand and Mr.\ M.\ Kusner of Saint-Gobain in Newbury, 
Ohio for interactions during the manufacture and characterization of
the CsI(Tl)crystals. We would also like to thank TRIUMF for
providing the personnel and infrastructure for the stand construction.
This work was supported in part by the U.S.\ Department of Energy
(Office of Energy Research, under Contract W-7405-ENG-36), the
National Science Foundation (Grants No. PHY-0100348 and PHY-0457219) and 
the NSF Major Research Instrumentation program (NSF-0116146), the Natural
Sciences and Engineering Research Council of Canada, and the Japanese
Grant-in-Aid for Scientific Research A12304014.

\end{document}